\documentclass[aps,prl,reprint,notitlepage,twocolumn,longbibliography]{revtex4-1}

\usepackage{graphicx}
\usepackage{amsmath}%
\usepackage{longtable}%
\usepackage{bm}%
\usepackage{color}

\def\avdis {\la\epsilon\ra}
\def\avdiss {\la\epsilon_s\ra}
\def\avdisd {\la\epsilon_d\ra}
\def\la {\langle}
\def\ra {\rangle}

\def\us {u_s}
\def\vus {\mathbf{u_s}}
\def\ud {u_d}
\def\vud {\mathbf{u_d}}

\newcommand{\rfig}[1]{Fig.~\ref{fig:#1}}

\begin{document}

\title{Solenoidal scaling laws for compressible mixing}


\author{John Panickacheril John}
\affiliation{Department of Aerospace Engineering, Texas A\&M University, College Station, Texas 77843, USA}
\author{Diego A. Donzis}
\email[]{donzis@tamu.edu}
\affiliation{Department of Aerospace Engineering, Texas A\&M University, College Station, Texas 77843, USA}
\author{Katepalli R. Sreenivasan}
\affiliation{Department of Mechanical and Aerospace Engineering, Department of Physics and Courant Institute of Mathematical Sciences, New York University, New York, New York 10012, USA}
\date{\today}

\begin{abstract}
Mixing of passive scalars in compressible turbulence does not obey the same classical Reynolds number scaling as its incompressible counterpart. We first show from a large database of direct numerical simulations that even the solenoidal part of the velocity field fails to follow the classical incompressible scaling when the forcing includes a substantial dilatational component. Though the dilatational effects on the flow remain significant, our main results are that both the solenoidal energy spectrum and the passive scalar spectrum scale assume incompressible forms, and that the scalar gradient aligns with the most compressive eigenvalue of the solenoidal part, provided that only the solenoidal components are used for scaling in a consistent manner. Minor modifications to this statement are also pointed out.
\end{abstract}

\maketitle
PACS numbers 47.27\\
                                                                                
A defining feature of turbulence is the ability to mix substances with orders of magnitude greater effectiveness than molecular mixing. The subject has been studied extensively \cite{sreeni2018} when the mixing agent is incompressible turbulence because it is a fundamentally important problem in its own right and a good paradigm for many practical circumstances. However, there are critically important applications from astrophysics to high-speed aerodynamics in which compressibility needs to be explicitly considered. Including compressibility renders inapplicable the Reynolds number scaling laws \cite{LELE1994,sarkarjfm1995} that are used extensively in incompressible turbulence. This paper shows one successful way of incorporating compressibility explicitly. We show by three examples that the standard incompressible laws work in the compressible case by rescaling the appropriate variables.

The initial attempt to include compressibility was through a suitably defined Mach number as an additional parameter. For the ideal case of homogeneous isotropic turbulence in a cubic box with periodic boundary conditions, this Mach number, $M_{t} =  u'/ \langle c \rangle$, where $u' = \langle u_{i}^{2}\rangle^{1/2}$, $u_i$
being the velocity in the Cartesian direction $i$, and the angular brackets indicate a suitable average. However, as has been pointed out by Ni \cite{ni2016}, $M_{t}$ is not adequate when the velocity field has a strong dilatational component. Indeed, DNS data with different types of large scale forcing, such as pure solenoidal forcing \cite{JD2016,DJ2013,WGWPRF2017}, homogeneous shear forcing \cite{chen2018spectra}, dilatational forcing \cite{WJWMCSXCCS2018,WYSXHC2013} and thermal forcing \cite{wangjfm2019}, have revealed that the dilatational flow field characteristics depend on the details of forcing, even for fixed $M_{t}$. Further progress has been made recently \cite{DJPNAS2019} by adding yet another parameter, namely $\delta = \ud'/\us'$, which is the ratio of root-mean-square (rms) dilatational to solenoidal velocity. These two components can be readily obtained for homogeneous compressible turbulence, by utilizing the Helmholtz decomposition of the velocity field, $\mathbf{u} = \vus + \vud,$ where the solenoidal part, $\vus$, represents vortical contribution and satisfies the incompressibility condition $\nabla \cdot \vus =0$. The dilatational part, $\vud$, represents the irrotational component and satisfies $\nabla \times \vud  = 0$. The improved physical understanding that arises from \cite{DJPNAS2019} can be used to assess the scaling of the passive scalars in compressible turbulence.

\begin{figure}[h]
\includegraphics[clip,trim=00 0 0 0,width=0.38\textwidth]{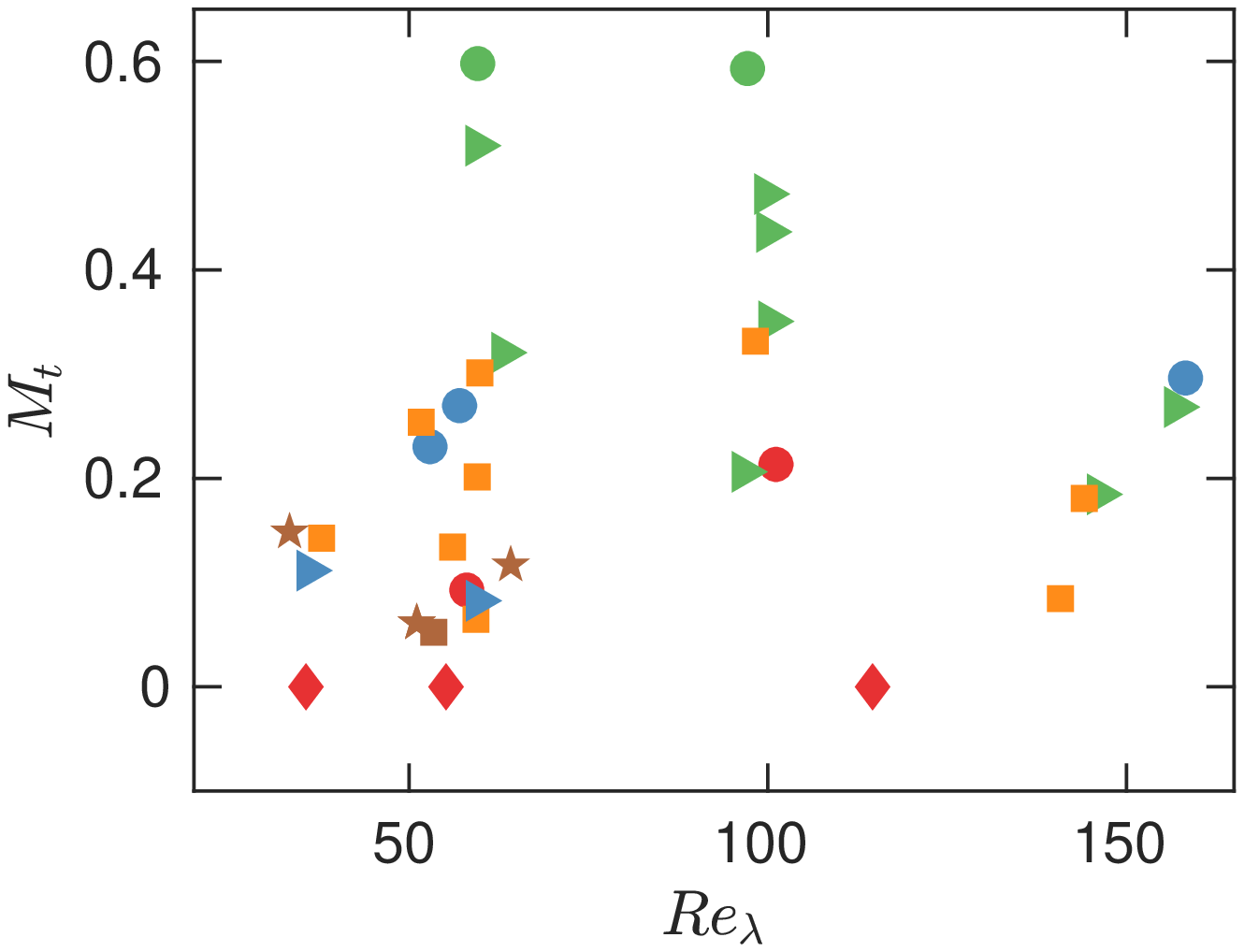}
\includegraphics[clip,trim=00 0 0 0,width=0.38\textwidth]{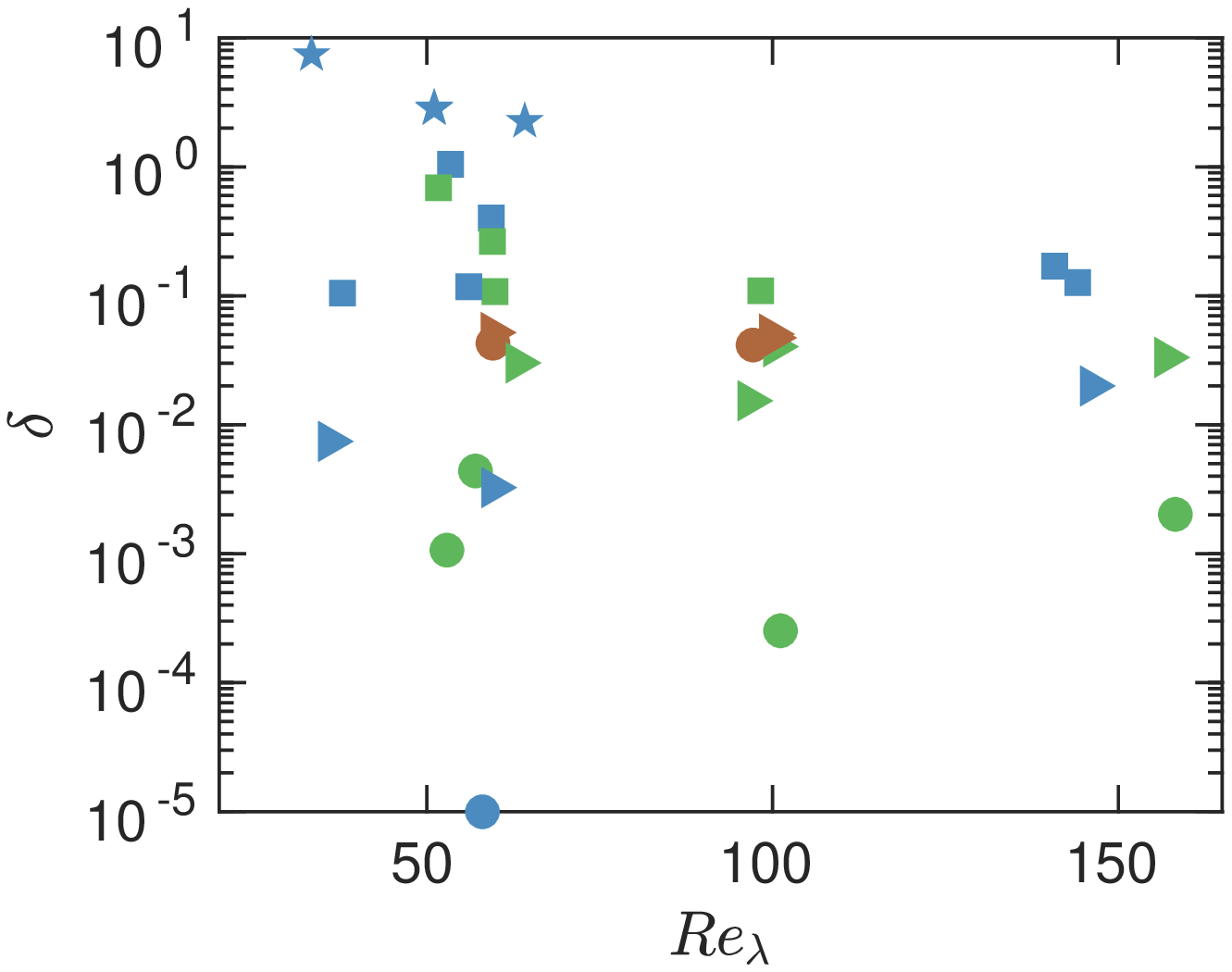}
\caption{\label{fig:1a} Parameter space of simulations for the scalar field.
(a) $M_{t}$-$R_\lambda$ plane.
Red: $\delta < 10^{-3}$, Blue: $10^{-3} <
\delta < 10^{-2}$, Green: $10^{-2} < \delta < 10^{-1}$, Orange: $10^{-1} <
\delta < 1$,  Brown:  $1 < \delta < 10$.
(b) $\delta$-$R_\lambda$ plane
Blue:
$M_{t} < 0.2$, Green: $0.2 < M_{t} < 0.4$, Brown:  $0.4 < \delta < 0.7$.
Symbols in all figures correspond to different percentages of dilatational
forcing, $\sigma$. diamonds: incompressible simulations; circles: $\sigma=
0$, triangles: $\sigma= 10-30$, squares: $\sigma= 30-65$, stars: $\sigma= 65-100$.}
\end{figure}

The data for the current work come from direct numerical simulations (DNS) of
compressible Navier-Stokes equations in a periodic box yielding homogeneous and
isotropic turbulence, and span the following conditions: 
the microscale Reynolds number $R_{\lambda} \equiv \la\rho\ra u^{'}
\lambda/\mu$, where $\la\rho\ra$ is the mean density,
$\lambda$ is the Taylor microscale and
$\mu$ the mean dynamic viscosity, ranges from 38 to 165;
the turbulent Mach number, $M_{t}$, varies between 0 and about 0.6;
the Schmidt number $Sc =
\mu/[\la\rho\ra D]$, where $D$ is the diffusivity of
the scalar, is unity.
The forcing
at low wavenumbers contains a strong dilatational component as well, with
$\delta$ ranging from 0 to 7.5. Figure \ref{fig:1a} shows the wide range of
compressibility conditions covered for the scalar field in the parameter spaces of $R_{\lambda}$,
 $\delta$ and $M_{t}$.

\begin{figure}[h]
\includegraphics[clip,trim=00 0 0 0,width=0.38\textwidth]{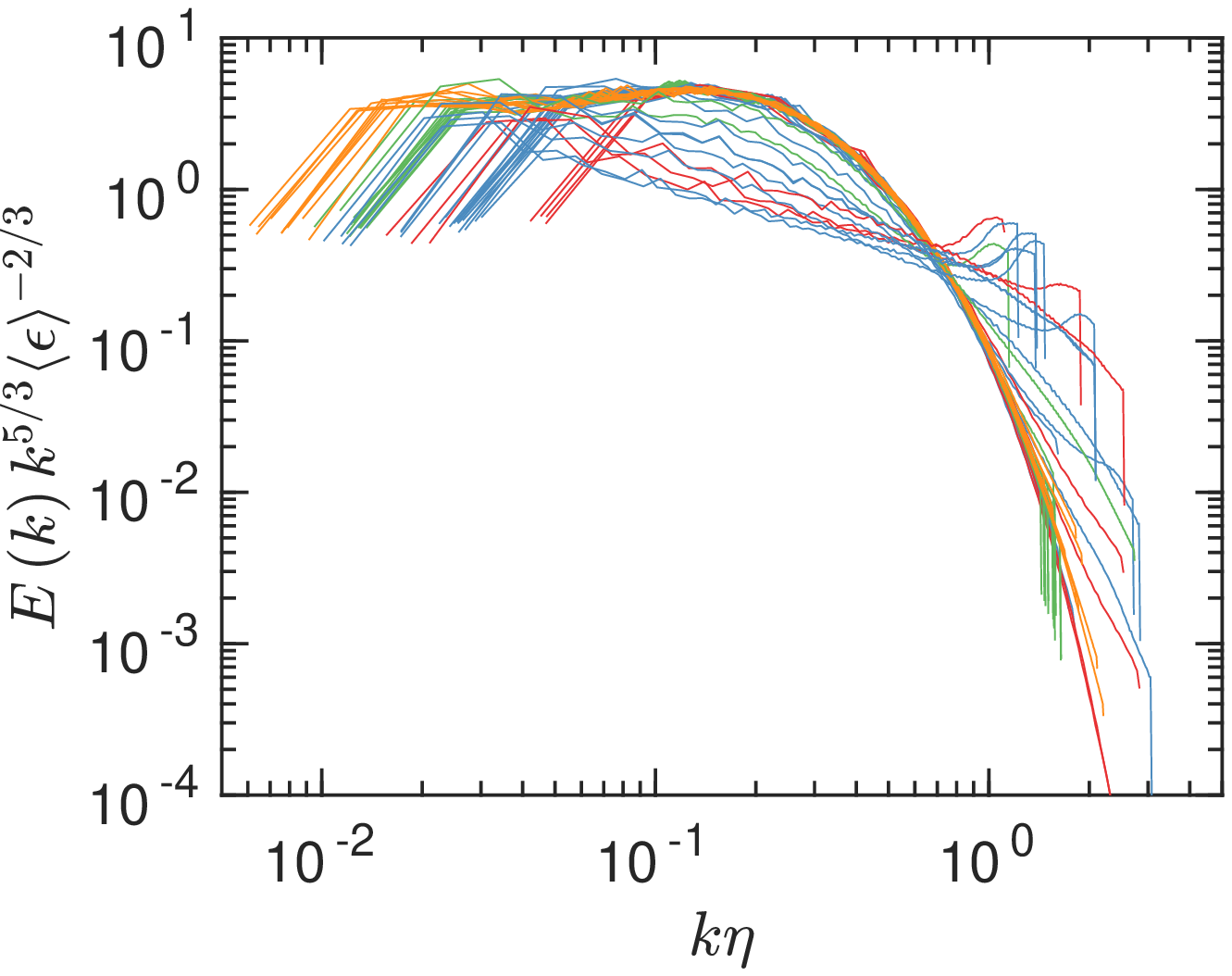}
\includegraphics[clip,trim=00 0 0 0,width=0.38\textwidth]{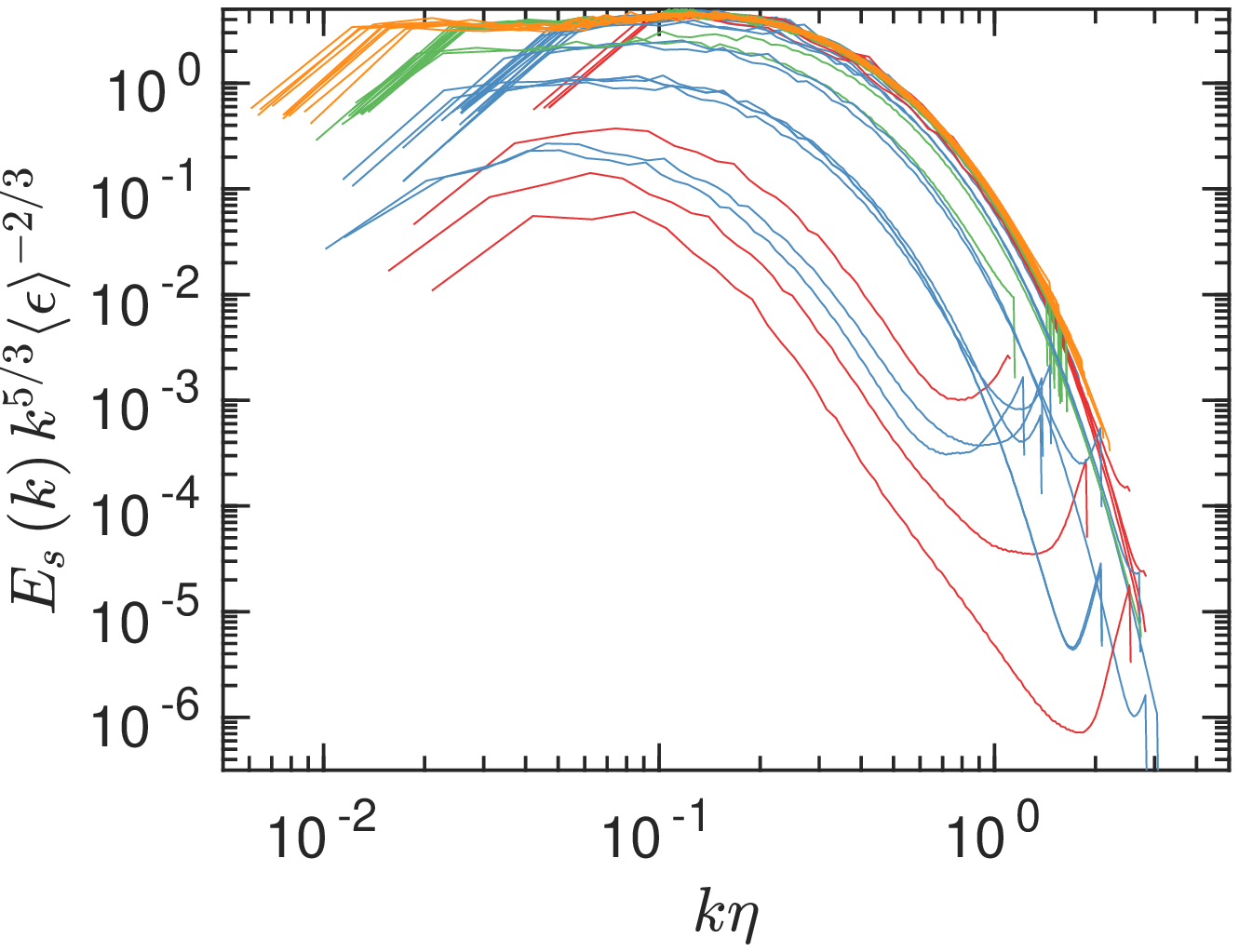}
\caption{\label{fig:a} 
(a) Kolmogorov-compensated total energy spectra using total energy dissipation
and Kolmogorov length scale 
($\eta = \left(\mu^3/\la\rho\ra^2\avdis \right)^{1/4}$) \cite{JD2016}.
(b) Kolmogorov-compensated solenoidal energy spectra using
total dissipation and the Kolmogorov length scale based on it. 
Here and in all figures to follow except the last, different colors correspond to different 
Reynolds numbers. Red: $R_{\lambda} < 40$, blue: $40 < R_{\lambda} < 75$,
green: $75 <R_{\lambda} < 115$, and orange: $115 < R_{\lambda} < 170$.
The velocity data of this figure and of Fig.\ 5(a) include larger set of conditions than shown for the scalar field in Fig.\ 1.
} 
\end{figure}

The first instance of the inadequacy of incompressible scaling is the energy spectrum which, according to \cite{K41}, follows the relation $E(k) = C \avdis^{2/3} k^{-5/3}$ in the inertial range, where $C$ is the Kolmogorov constant, $k$ is the wavenumber, and $\avdis$ is the mean total energy dissipation. The energy spectrum has the property that $\int_{0}^{\infty} E \left(k\right) dk = \langle u^{2} \rangle/2; \langle u^2 \rangle = \langle u_i u_i \rangle$. In Fig.\ \ref{fig:a}(a) we see that, unlike in incompressible turbulence, there is no collapse of spectral data when normalized according to \cite{K41}. This is not surprising: it has been pointed out already in theories \cite{ristorcelli1997,SC.book2008,SEHK1991} and simulations \cite{JD2016,WGWPRF2017,WJWMCSXCCS2018,WYSXHC2013} that the dilatational component of energy can take on a wide range of behaviors and can depart from the classical Kolmogorov scaling.

As an improvement, it has been suggested that the solenoidal part of the energy spectra $\left(\nabla \cdot \mathbf{u_{s}=0}\right)$ does scale according classical Kolmogorov scaling; the basis for this claim comes from solenoidally forced DNS \cite{JD2016,WGWPRF2017}. However, this result does not hold when the forcing has a strong dilatational component, as shown in Fig.\ \ref{fig:a}(b), where the Kolmogorov-compensated solenoidal energy spectra $E_s(k)$, defined such that $\int_{0}^{\infty} E_{s} \left(k\right) dk = \langle u_{s}^{2} \rangle/2,$ does not scale when the forcing includes a dilatational part. 

\begin{figure}[h]
\includegraphics[clip,trim=00 0 0 0,width=0.38\textwidth]{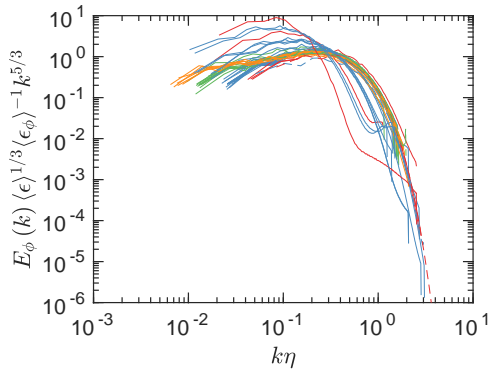}
\caption{\label{fig:a1}
The Obukhov-Corrsin compensated scalar spectra using total dissipation and Kolmogorov length scale based on the total energy dissipation. No scaling is observed. Dashed line is for the incompressible case.}  
\end{figure}

The second instance of this inadequacy is the scalar spectrum. In
incompressible turbulence, its behavior is reasonably well understood at the
phenomenological level
\cite{obuk1949,corrsin1951a,batch1959a,kraich68,WG2004,YXS2002,sreeni1996,sreeni2018}.
For unity Schmidt number, the appropriate normalization for the passive scalars
is the Obukhov-Corrsin normalization $E_{\phi}(k)  = C' \la\epsilon_{\phi}\ra
\avdis^{-1/3} k^{-5/3}$ where $E_{\phi}$ is defined such that
$\int_{0}^{\infty} E_{\phi} \left(k\right) dk =  \langle \phi^{2} \rangle/2$
and $\la\epsilon_{\phi}\ra$ is the mean scalar dissipation; $C'$ is the Obukhov-Corrsin
constant. In Fig.\ \ref{fig:a1}, we plot the Obukhov-Corrsin compensated scalar
spectra for all cases. There is no collapse of the data, and so compressibility
appears to have a first order effect on the scalar spectra.

\begin{figure}[h]
\includegraphics[clip,trim=00 0 0 0,width=0.4\textwidth]{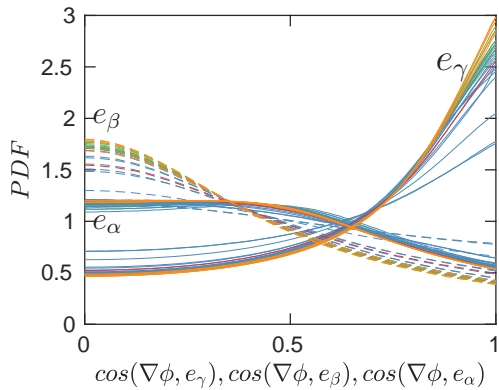}
\caption{\label{fig:b}  Alignment of scalar gradient $\left(\nabla \phi
\right)$ with the eigen-directions of the $S_{ij}$,
i.e.\ $e_{\gamma}, e_{\beta}, e_{\alpha}$ which correspond to 
the eigenvalues $\left(\gamma, \beta, \alpha \right)$ 
with $\gamma < \beta < \alpha$.} 
\begin{picture}(0,0)
\put(60,175){\Large{$e_{\gamma}$}}
\put(-70,155){\large{$e_{\beta}$}}
\put(-70,115){\large{$e_{\alpha}$}}
\end{picture}
\end{figure}

As a third quantity, consider the alignment of the scalar gradient with the directions of the eigenvectors of the strain field. In incompressible turbulence, the turbulent velocity field plays an important role in the stirring action of passive scalars where the different isosurfaces of the scalars are brought together \cite{warhaft2000,dimo05,sreeni2018}. This stirring action results in high scalar gradients across the flow field, ultimately enabling molecular diffusion to act. Batchelor's theory \cite{batch1959a}, initially proposed for large Schmidt numbers, shows that the scalar gradient aligns itself with the most compressive eigenvalue. DNS studies \cite{DSY2010} have shown that this aspect of the theory is valid, perhaps surprisingly, even for Schmidt numbers of order unity; see also Vedula \textit{et al.} \cite{VYF2001}. Danish \textit{et al.} \cite{DSG2016} studied this alignment for decaying compressible turbulence and found that the topology and alignment were universal for a range of Reynolds and Mach numbers, though their studies were confined to a narrow range of initial $M_{t}$ $\left(0.50-0.70\right)$ and $R_{\lambda}\left(18 - 24\right)$.  For the wider range of compressible turbulent states considered here, in terms of $R_{\lambda}$, $M_{t}$ and $\delta$, Fig.\ \ref{fig:b} shows that the scalar gradient, $\nabla \phi = \partial \phi /\partial x_{i}$, does not align uniquely with the symmetric part of the velocity gradient tensor, $S_{ij}$, where 
$$S_{ij} = 
\frac{1}{2}\left(
\frac{\partial u_{i}}{\partial x_{j}} 
+ \frac{\partial u_{j}}{\partial x_{i}}
\right) . $$ 
The eigenvectors of this tensor, called here 
$e_{\alpha}$, $e_{\beta}$, and $e_{\gamma}$, 
correspond respectively to the maximum, intermediate and minimum eigenvalues with
$\alpha > \beta > \gamma$; incompressible turbulence is constrained by
$\alpha + \beta + \gamma = 0$. The previous observations by Blaisdell \textit{et al.} \cite{BMR1994}, and more recently by Ni  \cite{ni2016}, that contributions from the dilatational field to the scalar flux are negligible compared to the solenoidal part alone, correspond to a narrow range of conditions.

The discussion so far makes it clear that even the spectrum for just the solenoidal part of the velocity field does not satisfy the incompressibility scaling laws if we consider forcing with a dilatational component (see Fig.\ 1). Existing work \cite{JD2016,WGWPRF2017,wangpf2011effect,WGWinterPRF2017} which makes this claim concerns the velocity field under solenoidal forcing and decaying turbulence. 

\begin{figure}[h]
\includegraphics[clip,trim=00 0 0 0,width=0.45\textwidth]{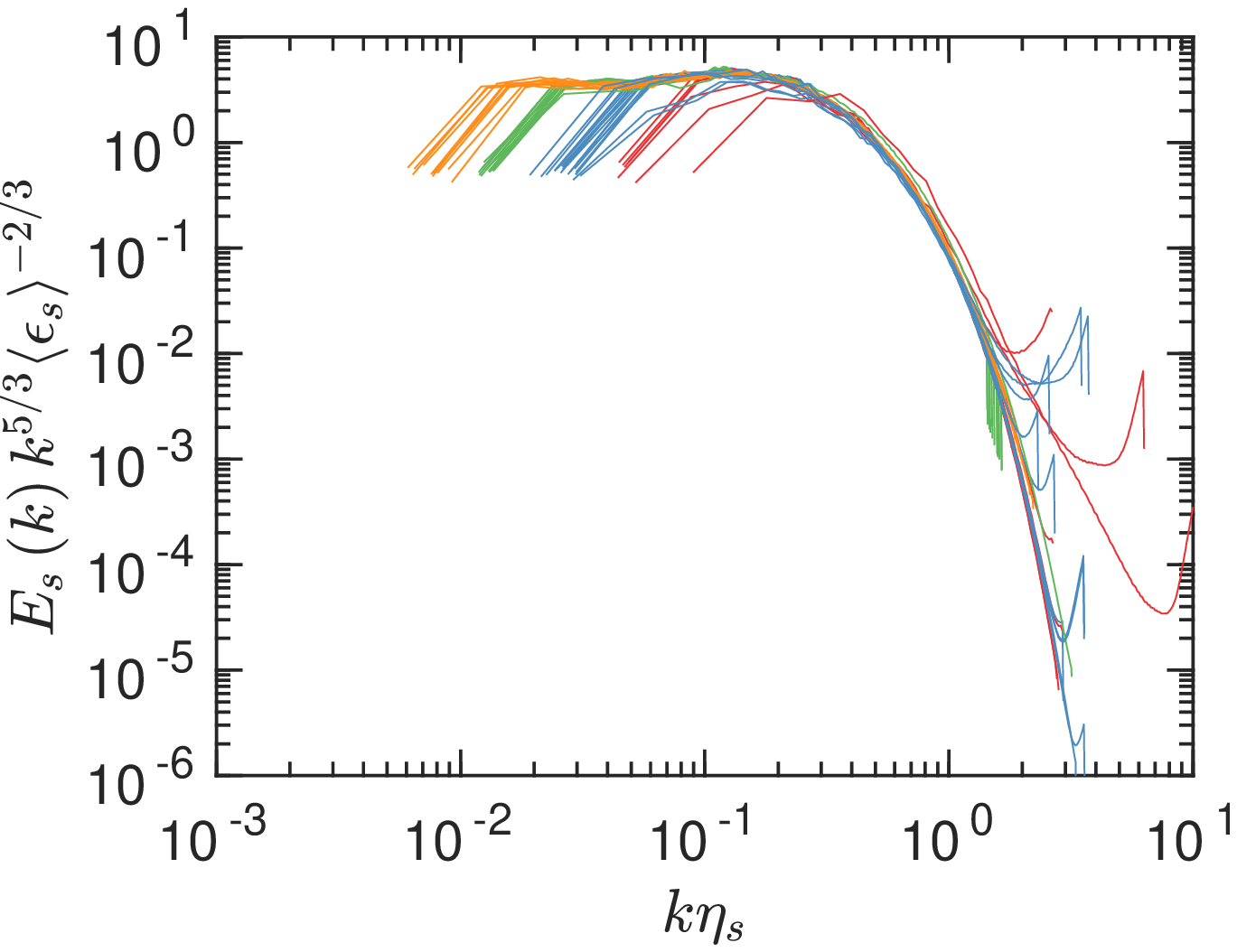}
\includegraphics[clip,trim=00 0 0 0,width=0.45\textwidth]{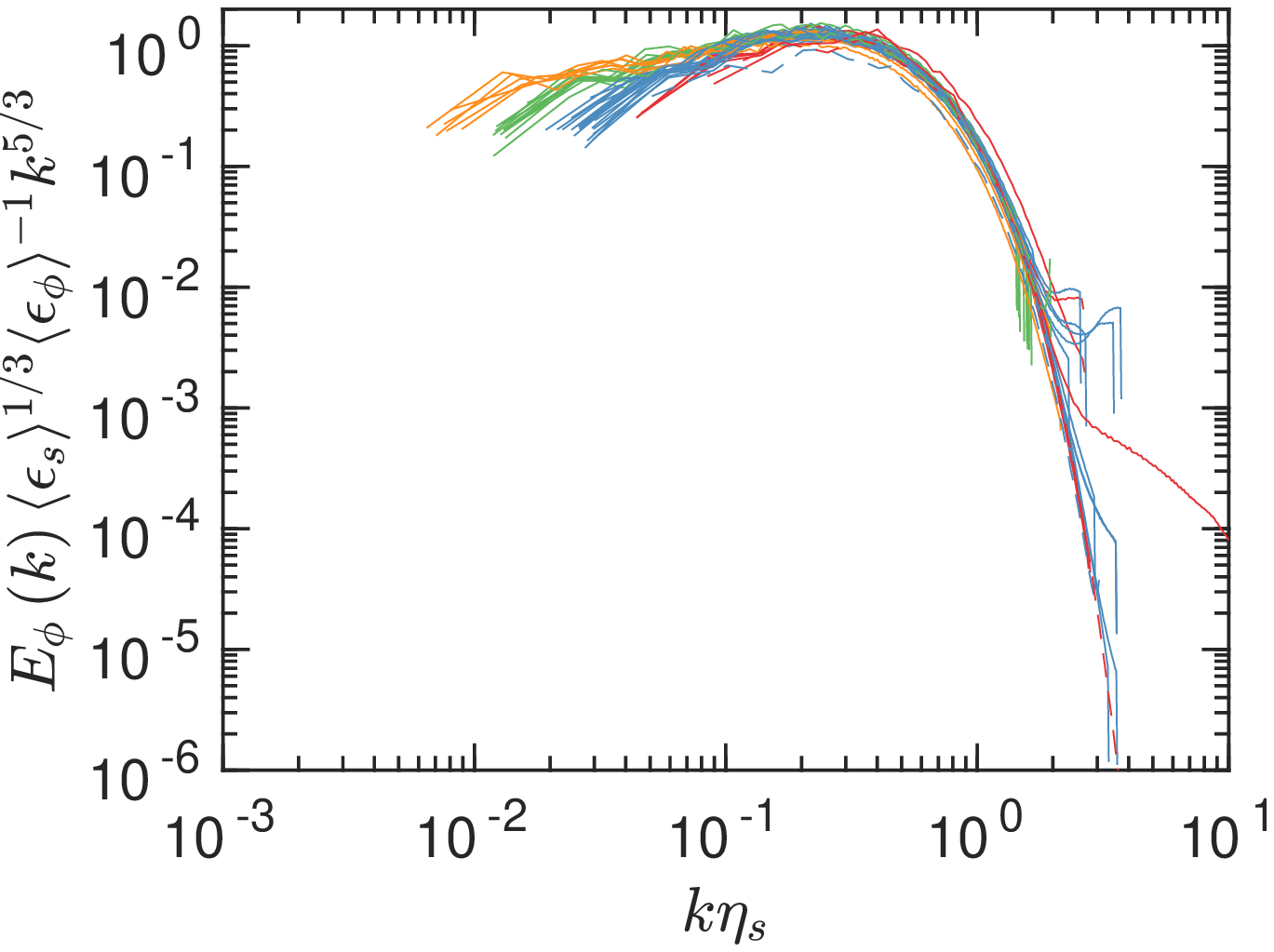}
\caption{\label{fig:c} Kolmogorov-compensated solenoidal
energy spectra (a) and 
Obukhov-Corrsin compensated scalar spectra (b) using
solenoidal dissipation,$\avdiss$ and solenoidal Kolmogorov
length scale, $\eta_{s}$.  Dashed line in the bottom figure is for the incompressible case. }
\end{figure}

We now propose the following paradigm. Similar to the velocity field one can
decompose the dissipation into solenoidal and dilatational contributions as
$\avdis = \avdiss + \avdisd,$ where $\avdiss = \langle \mu \omega_{i}
\omega_{i} \rangle$, $\boldsymbol{\omega}$ being the vorticity of the fluid
motion, and $\avdisd  = (4/3) \langle \mu \left( \partial u_{i} / \partial
x_{i}\right)^{2}\rangle$ are the solenoidal and dilatational parts,
respectively. Indeed, under solenoidal forcing conditions when $\delta \ll 1$
and $\avdiss \approx \avdis$, we do not expect significant departures in the
scaling of the solenoidal energy spectra. 
However, under general conditions of mixed solenoidal-dilatational forcing where $\delta$ can vary by orders of magnitude, one may expect using solenoidal variables in the compensation of the solenoidal spectra would yield
better collapse. Indeed, \rfig{c}(a) shows the excellent collapse of the Kolmogorov-compensated solenoidal energy spectra when both velocity and the dissipation pertain solely to the solenoidal variables. The solenoidal Kolmogorov length scale is defined \cite{JD2016} as $\eta_s = (\la\mu\ra^3/\la\rho\ra^2\avdiss)^{1/4}$.

In Fig.\ \ref{fig:c}(b), we plot the
Obukhov-Corrsin compensated scalar spectrum using just the solenoidal part of
the velocity field. A robust collapse
occurs for scalar spectra under a wide range of conditions and the spectra look
similar to the incompressible case. 
This suggests that even at really high
levels of dilatational content in the flow field, the interaction between the
passive scalars and solenoidal velocity field is universal. 
The implication is that 
the cascade process in which the large scales of the passive scalar are broken
down to smaller scales is independent of compressibility. Thus classical
scaling laws, when modified by suitable rescaling, obey the
same incompressible turbulence models in highly compressible flows, even when
the dilatational part is quite strong.  

We now come to the orientation of the scalar gradient with respect to the velocity strain field. Following the observations above, we assess the effect of the solenoidal component of the tensor, $S^s_{ij}$. In particular, we examine the statistics of the normalized eigenvalues
($\beta_{s}$) \cite{VYF2001} given by
$\widehat{\beta}_{s} =  {\sqrt{6} \beta_{s}} \big/
{\sqrt{\alpha_{s}^{2} + \beta_{s}^{2} + \gamma_{s}^{2}}}$,
such that $-1\le \widehat{\beta}_s\le 1$.

\begin{figure}[h]
\includegraphics[clip,trim=00 0 0 0,width=0.45\textwidth]{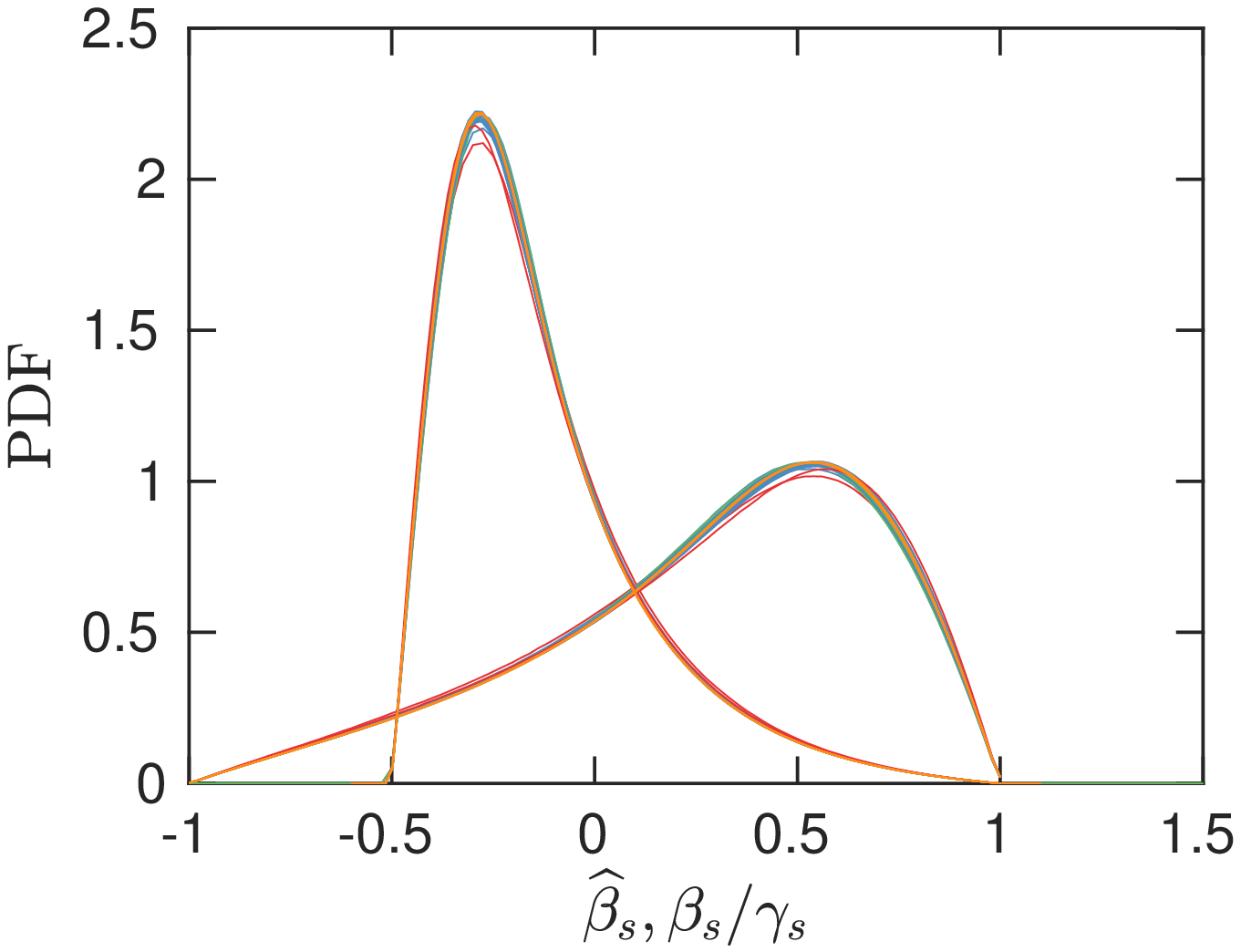}
\includegraphics[clip,trim=00 0 0 0,width=0.45\textwidth]{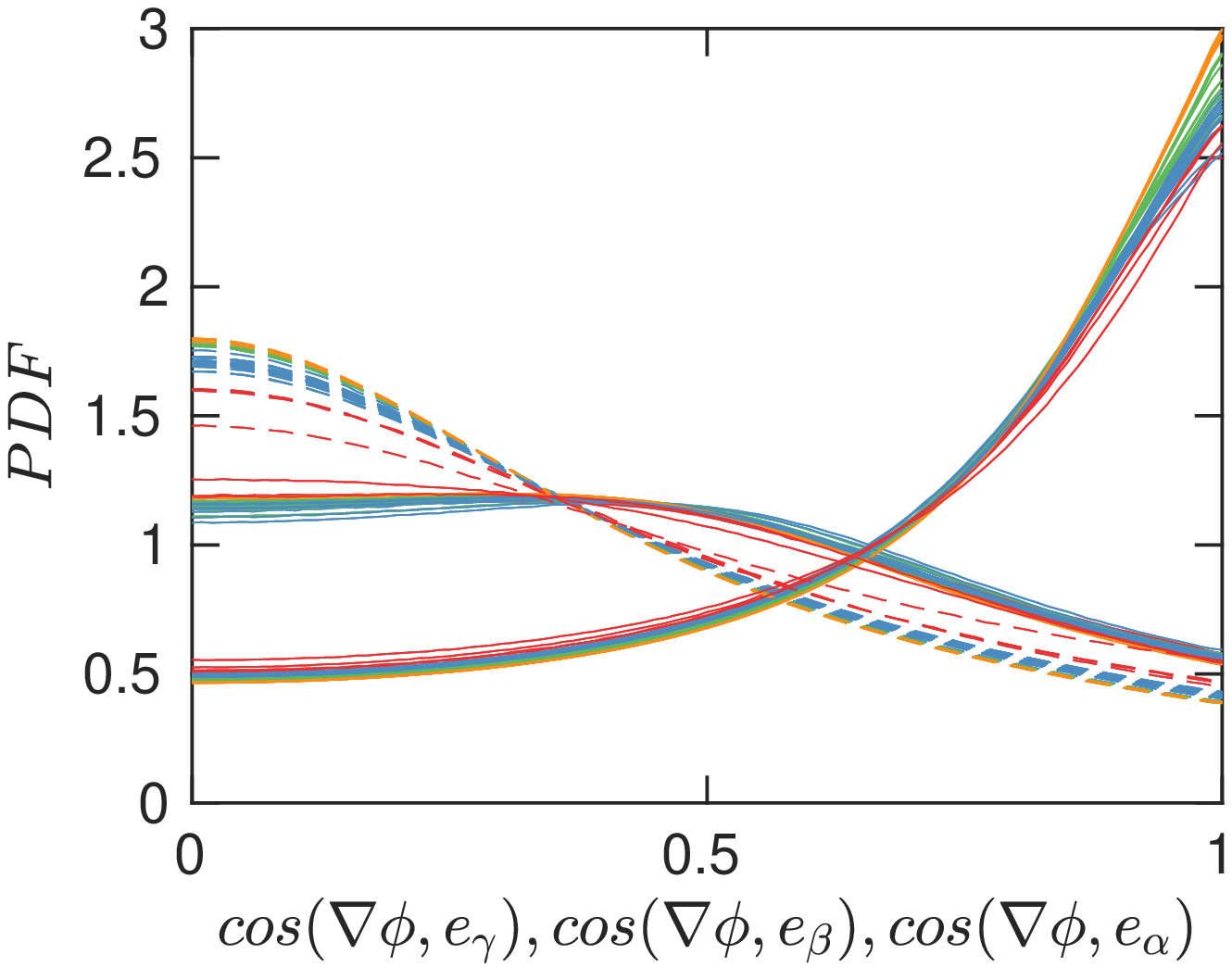}
\caption{\label{fig:d} 
(a) Normalized eigenvalues of the solenoidal symmetric velocity gradient tensor,
$S^s_{ij}$:
 \romannumeral 1 )
 $\widehat{\beta}_s=\sqrt{6}\beta_{s}/\sqrt{\alpha_{s}^{2} + \beta_{s}^{2} + \gamma_{s}^{2}}$; 
\romannumeral 2 )
 $\beta_{s} / \gamma_{s}$.
(b) Alignment of scalar gradient $\left(\nabla \phi \right)$ 
with $e_{\gamma}, e_{\beta}, e_{\alpha}$,
the eigenvectors of $S^{s}_{ij}$.
} 
\begin{picture}(0,0)
\put(75,220){\Large{$e_{\gamma}$}}
\put(-75,180){\Large{$e_{\beta}$}}
\put(-75,137){\Large{$e_{\alpha}$}}
\put(55,315){\Large{\romannumeral 1}}
\put(-55,350){\Large{\romannumeral 2}}
\end{picture}
\end{figure}

In Fig.\ \ref{fig:d}(a) is plotted the probability density function (PDF)
of $\widehat{\beta}_s$ for a wide range of compressibility conditions. Excellent collapse is observed (curve (i)), indicating that the ratio of
the PDF  
of the eigenvalues is unaffected by compressibility. 
Similar universal behavior is observed for 
the ratio of $\beta_{s} /\gamma_{s}$ shown as curve (ii) in the same figure. 
We also note that the maximum probable
value of $\beta_{s} / \gamma_{s}$ is approximately $0.28$ which corresponds to the
ratio of $\gamma_{s} / \beta_{s} = 3.7$, close to the situation
suggested for incompressible turbulence \cite{AKKG1987} and consistent with
results for solenoidal forcing \cite{WSWX+2012}. This feature suggests that,
while compressibility may change the solenoidal field itself, it 
does not alter its mixing capability 
and would remain as efficient as incompressible
turbulence.      
  
\begin{figure}[h]
\includegraphics[clip,trim=00 0 0 0,width=0.45\textwidth]{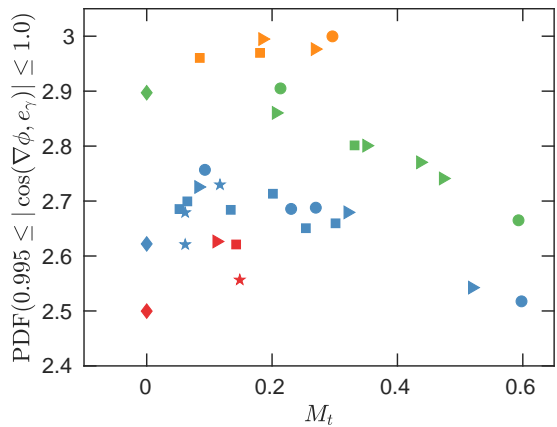}
\caption{\label{fig:e} 
(a) 
Probability of scalar gradient $\left(\nabla \phi\right)$ 
being perfectly aligned with the eigendirection $e_{\gamma}$
corresponding to the most compressive eigenvalue.
Symbols in the figure correspond to different percentages of dilatational
forcing, $\sigma$: 
incompressible simulations (diamons), 
$\sigma=0$ (circles), 10-30 (triangles), 30-65 (squares), and 65-100 (stars).
} 
\begin{picture}(0,0)
\end{picture}
\end{figure}

Figure \ref{fig:d}(b) plots the alignment of the scalar gradient with
the solenoidal frame of reference. One finds that the behavior of 
the scalar gradient
is very similar to that of incompressible turbulence \cite{VYF2001}, 
with a high probability
for the scalar gradient to align with the most compressive
direction. There are, however, some weak compressibility effects. To
understand them qualitatively, we show in Fig.\ \ref{fig:e} the PDF values
for $\cos\left(\nabla \phi,e_{\gamma}\right)\in [0.995,1]$---that is, when the two
vectors are almost perfectly aligned---as a function of turbulent Mach number,
$M_{t}$. The figure shows that $R_{\lambda}$ is the major effect, though
a weaker decreasing trend with $M_{t}$ is also seen.
In order to completely understand
compressible turbulent mixing, one has to include these secondary
compressibility effects on the fine scale structure of turbulence.

In summary, using high fidelity DNS data, we have shown that the interaction between passive scalar and solenoidal velocity field is universal under a wide range of compressibility conditions, for both the velocity and the scalar field, if both the velocity field and the energy dissipation are taken from the solenoidal part of the velocity.  

%


\end{document}